\documentclass[pdflatex,sn-mathphys-num]{sn-jnl}

\usepackage{hyperref}
\usepackage{graphicx}%
\usepackage{multirow}%
\usepackage{amsmath,amssymb,amsfonts}%
\usepackage{amsthm}%
\usepackage{mathrsfs}%
\usepackage[title]{appendix}%
\usepackage{xcolor}%
\usepackage{textcomp}%
\usepackage{manyfoot}%
\usepackage{booktabs}%
\usepackage{algorithm}%
\usepackage{algorithmicx}%
\usepackage{algpseudocode}%
\usepackage{listings}%
\usepackage{marginnote} 

\theoremstyle{thmstyleone}%
\theoremstyle{thmstyletwo}%

\theoremstyle{thmstylethree}%

\usepackage{xcolor}

\begin{document}

\title{Axion-like Particle Search with a Light-Shining-Through-Walls Setup at a $\gamma$-$\gamma$ Collider }


\author[1]{Zi-Yao Yan}

\author*[1]{Jie Feng}
\email{fengj77@mail.sysu.edu.cn}

\affil*[1]{\orgdiv{School of Science}, \orgname{Shenzhen Campus of Sun Yat-sen University}, \orgaddress{\street{NO.66 GongChangLu}, \city{Shenzhen}, \postcode{518107}, \state{Guangdong}, \country{China}}}


\date{}


\abstract{In this work, we have explored a practical extension of the conventional light-shining-through-walls technique by making direct use of the high-intensity $\gamma$-ray beam available at a $\gamma$-$\gamma$ collider. The energetic and highly collimated photon flux produced via inverse Compton scattering naturally provides an efficient axion-like particles production stage, while the addition of a regeneration region downstream enables a complete LSW configuration without introducing new experimental complexities. This approach therefore represents an experimentally simple and infrastructure-compatible method for enhancing laboratory sensitivity to axion-like particles (ALPs). 
Under conservative assumptions, we find that one year of operation can probe ALP-photon coupling down to
$g_{a\gamma\gamma}\simeq 3.84\times 10^{-5} \,\mathrm{GeV^{-1}}$ for $m_a\lesssim 0.1 \,\mathrm{eV}$ when an additional magnetic region is included upstream of the beam dump, improving upon previous laboratory LSW sensitivity by up to a factor of 4. }

\maketitle

\section{Introduction}\label{sec1}

Astronomical observations support the existence of non-baryonic dark matter (DM). This is exemplified by the flat rotation curves of galaxies, where stars orbit at constant high velocities far from the galactic center, indicating unseen mass \cite{McGaugh:2016leg,Spergel:2015noa}. At a large-scale, the distribution and growth of cosmic structures, from clusters to superclusters and vast voids, are best explained by the gravitational influence of cold DM \cite{Blumenthal:1984bp}. Gravitational lensing is a technique in astrophysics and cosmology to map DM distributions \cite{Wang:2010ma,Massey:2010hh}. Further astronomical evidence for DM comes from observations of galaxy clusters, such as the prominent Bullet Cluster. These observations utilize $X$-ray emissions to identify baryonic gas and gravitational lensing to infer the total mass distribution. And it exhibits a clear spatial separation between the $X$-ray emitting gas, from collisions of ordinary matter, and the total mass distribution, predominated by DM \cite{Shan:2010rz,Markevitch:2003at}. This decoupling provides strong evidence for the non-collisional nature of DM, as its components pass through each other with minimal interaction, unlike the gas.  Following the astronomical confirmation of DM's existence, cosmological observations have been instrumental in quantifying its abundance within the universe. According to measurements of cosmic microwave background anisotropies, DM accounts for approximately 27\% of the universe's total energy density \cite{Planck:2018vyg,Bertone:2004pz,Planck:2015fie}. While these astrophysical and cosmological phenomena confirm DM’s existence, its properties remain unknown.

Physicists study DM particles in three primary ways: i) measurement of annihilation or decay products (e.g. positrons or antiprotons by AMS-02 \cite{AMS:2014bun,AMS:2013fma,AMS:2021nhj,AMS:2016oqu,AMS:2019iwo,AMS:2019rhg} and PAMELA \cite{PAMELA:2008gwm,PAMELA:2010kea}, $\gamma$ rays by Fermi-LAT \cite{Abdo:2010nc,Abazajian:2012pn,Fermi-LAT:2015att}, or neutrinos via IceCube \cite{IceCube:2018cha,Yan:2023hpt}); 
ii) measurement of nuclear recoils in shielded experiments (e.g. LUX-ZEPLIN \cite{LZ:2022lsv}, XENONnT \cite{XENON:2024hup}, or PANDAX \cite{PandaX-II:2016vec}); 
iii) production of new particles in colliders (e.g. ATLAS \cite{Belyaev:2020wok}, CMS \cite{deCosa:2015smf,Berlin:2018jbm}). 
In this context, light-shining-through-walls (LSW) experiments can be viewed as belonging to the “production of new particles” category, but realized in a purely laboratory environment rather than a high-energy collider, making them particularly suitable for probing very light and weakly coupled degrees of freedom such as axion-like particles (ALPs).

ALPs are compelling theoretical candidates that offer potential solutions to several unresolved physical problems. First, the quantum chromodynamics (QCD) axion arises from the spontaneous breaking of a global U(1) Peccei-Quinn symmetry and solves the strong CP problem \cite{Peccei:1977hh,Weinberg:1977ma,Wilczek:1977pj,Berezhiani:1989fp}. More generally, pseudo-Goldstone bosons from broken global symmetries are often referred to as ALPs, which can span a wide mass range and couple weakly to Standard-Model fields. Second, the mass of ALPs spans a wide range, from ultralight $\left(<  10^{-14} \,\mathrm{eV}\right)$ to GeV scale, and they couple very weakly to Standard Model fields \cite{Dine:1982ah,Abbott:1982af,Preskill:1982cy,Dine:1981rt,Sakharov:1996xg}. This makes ALPs well-motivated candidates for cold dark matter in certain regions of parameter space. Their ultralight mass and weak interactions align well with the phenomenology of cold DM \cite{Choi:2020rgn}.

Constraints on the ALPs parameter space arise from astrophysical observations and a variety of laboratory searches. Astrophysical bounds, such as those from excess energy loss in stellar environments (e.g., SN 1987A \cite{Bar:2019ifz}), suffer from uncertainties in stellar  modeling.  Laboratory experiments can be broadly categorized into haloscopes and photon regeneration experiments. Haloscopes, such as ADMX \cite{ADMX:2025vom}, probe ALPs dark matter and therefore depend on assumptions about the local halo dark-matter density. Photon-regeneration experiments include both helioscopes, which rely on ALPs produced in the Sun (e.g., CAST \cite{CAST:2024eil,CAST:2007jps,CAST:2017uph,CAST:2004gzq}) and are consequently model-dependent, and purely laboratory-based regeneration experiments (e.g., NOMAD \cite{NOMAD:2000usb}, Inada \cite{Inada:2016jzh}, EuXFEL \cite{Halliday:2024lca}), which avoid astrophysical systematics by generating and detecting ALPs under controlled experimental conditions \cite{Jaeckel:2006xm}. These differences highlight the need for complementary detection strategies to robustly constrain the ALP–photon coupling 
($g_{a\gamma\gamma}$).

Within the class of photon-regeneration experiments, LSW setups constitute a particularly clean and model-independent method. The original LSW concept was introduced in Ref.~\cite{Sikivie:1983ip}. Modern laboratory implementations can be divided into (i) magnetic field LSW experiments \cite{Inada:2016jzh,Kling:2022ehv,Balkin:2021jdr} and (ii) crystal-based searches that exploit the strong electric fields in crystals \cite{Yamaji:2018ufo,Halliday:2024lca,NA64:2020qwq}. In all cases, photons convert to ALPs in the first conversion region, traverse an optical barrier, and then reconvert into detectable photons in a second regeneration region. Their sensitivity scales with the field strength, the available photon flux, and the accessible ALP mass range.
However, in the sub-eV to eV mass region, a large gap remains between current laboratory LSW sensitivity, which is at the level of $10^{-4}\mathrm{GeV}^{-1}$, and the QCD axion target, which is around $10^{-9}\mathrm{GeV}^{-1}$ near $m_a \sim 1$ eV. The novelty of the present proposal is that it identifies a practical way to push laboratory LSW searches toward this gap by combining a high-intensity beam with much higher photon energy than in conventional LSW setups.

In a future $\gamma$-$\gamma$ collider, a photon-regeneration experiment could be set up and probe $g_{a\gamma\gamma}$ down to $10^{-5}\,\mathrm{GeV^{-1}}$ when ALPs mass is close to $1\,\mathrm{eV}$, challenging existing astrophysical limits while providing model-independent constraints.

This paper is organized as follows. Section~\ref{Sec::method} outlines the experimental requirements and detection methodology. Section~\ref{Sec::result} evaluates the theoretical framework and projected sensitivity. Section~\ref{Sec::Summary} summarizes the implications for our DM search.

\section{Methodology}
\label{Sec::method}
\subsection{The \texorpdfstring{$\gamma$-$\gamma$}{gamma-gamma} Collider}
\begin{figure}
        \centering
        \includegraphics[width=1\linewidth]{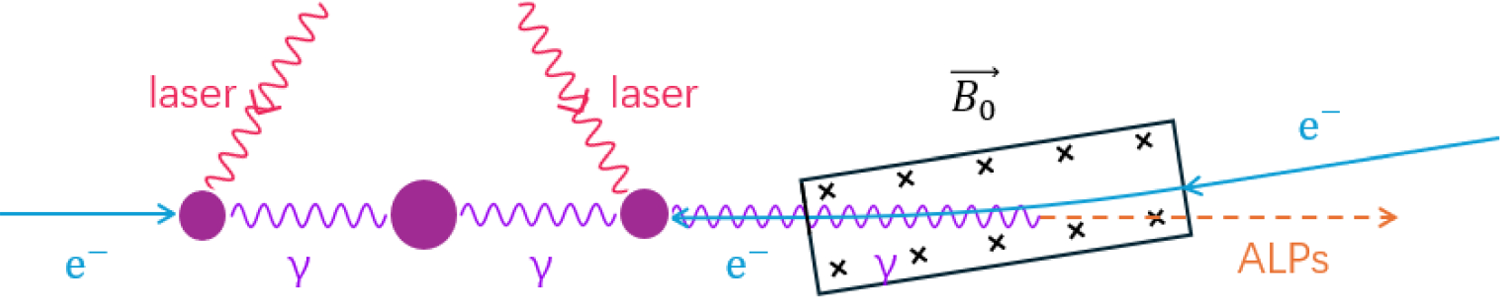}
        \caption{ Schematic of the $\gamma$-$\gamma$ collider. The $200\,\mathrm{MeV}$ electrons will be deflected by a deflected magnetic field $\vec{B_0}$ to adjust their direction. The high-energy $\gamma$-rays are generated through the inverse Compton scatterings between the electrons and low-energy photons \cite{zhou2025estimationinversecomptonscattering}. The high-energy $\gamma$-ray will pass through the magnetic field $\vec{B_0}$ and be converted into ALPs by the Primakoff effect.}
        \label{fig:gamma-collider}
 \end{figure}
The $\gamma$-$\gamma$ collider is a new experimental device that utilizes inverse Compton scattering of a laser and a high-energy electron beam to generate a $\gamma-ray$ with energy in the $\mathrm{keV}-\mathrm{MeV}$ range, extremely bright and highly collimated \cite{zhou2025estimationinversecomptonscattering}, as shown in Fig.~\ref{fig:gamma-collider}.
The $\gamma$-$\gamma$ collider employs two symmetric laser–-electron interaction points placed on opposite sides of the setup. At each point, inverse Compton scattering occurs between a $200\,\mathrm{MeV}$ electron beam and a high-power laser pulse with a $1054\,\mathrm{nm}$ wavelength, $2\,\mathrm{J}$ pulse energy, $10\,\mathrm{\mu}m$ waist, $1\,\mathrm{ps}$ duration, and a $50\,\mathrm{Hz}$ repetition rate. Each interaction point generates a highly directional high-energy $\gamma$-ray beam, and the two beams are directed to collide in the central region. Because the 
$\gamma$-$\gamma$ scattering cross-section is extremely small, the beams remain essentially undepleted after the collision and can subsequently be used for ALPs searches. The electrons are magnetically deflected away from the photon path, while the outgoing $\gamma$-ray beam propagates into the magnetic-field region where ALPs production via the Primakoff effect may occur. This configuration provides a compact, fully laboratory-controlled platform for probing ALP-photon coupling.

\subsection{ALPs Detection Setup}

\begin{figure}
    \centering
    \includegraphics[width=1\linewidth]{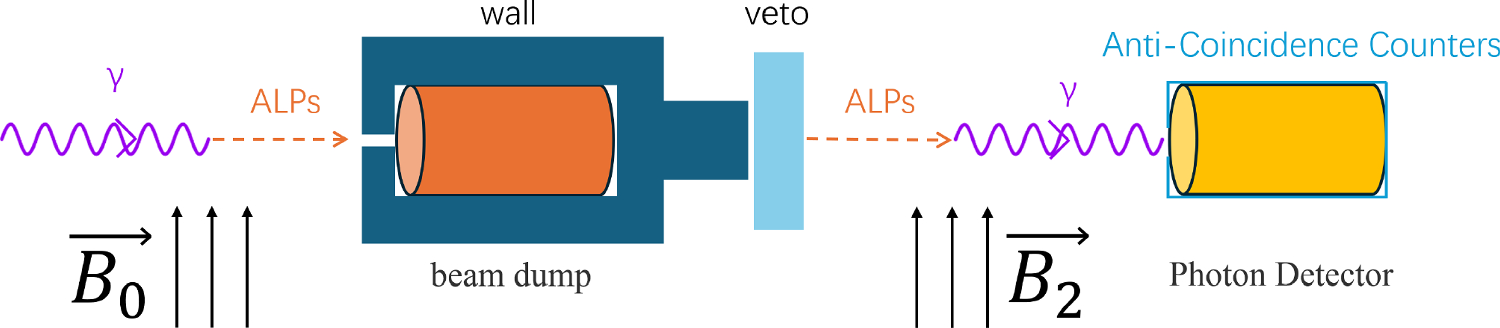}

    \caption{
Schematic layout of the light-shining-through-walls setup implemented at the
$\gamma$-$\gamma$ collider. High-energy $\gamma$ rays produced via inverse
Compton scattering first traverse the magnetic-field region $\vec{B}_0$
(and optionally an additional region $\vec{B}_1$), where a fraction of the
photons convert into axion-like particles (ALPs) through the Primakoff effect. The primary $\gamma$-ray beam is removed by a beam dump system before reaching the wall, so that the wall is not irradiated by high-energy photons. The wall serves only to absorb residual or scattered photons, while ALPs pass through unimpeded. Downstream of the wall, a scintillator is placed to be used as a veto of leakage photons. A second magnetic-field region $\vec{B}_2$ enables photon regeneration via the inverse Primakoff effect. The Photon Detector records regenerated photons behind the wall.}

    \label{fig:experimental_setup}
\end{figure}

\begin{figure}
    \centering
    \includegraphics[width=1\linewidth]{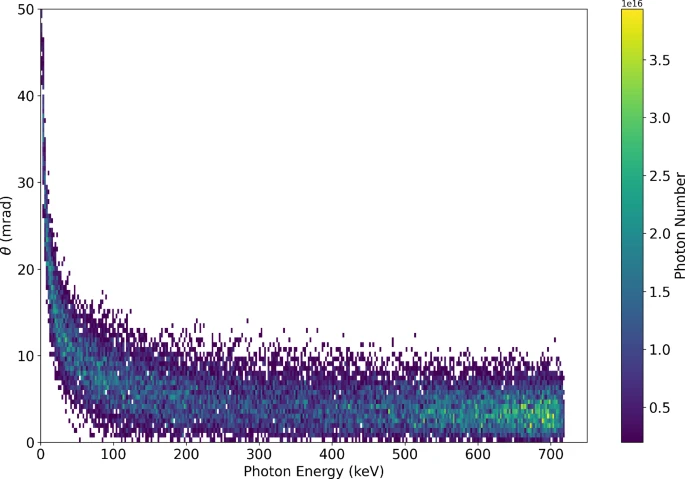}

    \caption{
A year of simulated photon samples from inverse Compton scattering in the [$\theta$-Energy] plane. The plot illustrates the kinematic correlation between photon energy and scattering angle $\theta$, showing a continuous broadband spectrum that extends to a Compton edge of approximately 720 keV \cite{zhou2025estimationinversecomptonscattering}.
}

    \label{fig:Spectrum}
\end{figure}

Our experimental configuration, shown in Fig.~\ref{fig:experimental_setup}, is built on a 
$\gamma$-$\gamma$ collider capable of producing high-energy 
$\gamma$-rays at a rate of 
$1.45\times10^{12}\,\mathrm{Hz}$ \cite{zhou2025estimationinversecomptonscattering}. 
The energy and angular distributions of these $\gamma$-rays are presented in Fig.~\ref{fig:Spectrum} \cite{zhou2025estimationinversecomptonscattering}. 
After production, the photons naturally traverse the collider’s deflection magnetic field $|\vec{B}_{0}|=0.105\,\mathrm{T}$ with an effective length 
$L_0=1\,\mathrm{m}$, where a fraction of them convert into ALPs via the Primakoff effect. In order to increase the probability of photon-ALPs conversion, we can add an additional magnetic field $|\vec{B}_{1}|=5\,\mathrm{T}$ with an effective length 
$L_1=0.2\,\mathrm{m}$ in the collider. The photon polarization is tunable to be parallel with the magnetic field.

In the experimental setup, the beam dump is a lead cylinder of diameter $5\,\mathrm{cm}$ and thickness $17\,\mathrm{cm}$, placed along the photon propagation path to intercept the high-energy $\gamma$-rays. The required thickness was determined by a dedicated Geant4~11.4.0 simulation using the full beam phase-space from the $\gamma$-$\gamma$ collider ($1.45\times10^{12}\,\mathrm{Hz}$, one year of operation, $4.67\times10^{19}$ real photons). The simulation employs the \texttt{G4EmStandardPhysics\_option4} physics list (Livermore/Penelope electromagnetic models) and tracks all secondary particles, i.e., bremsstrahlung photons, Compton-scattered $\gamma$-rays, and Pb fluorescence X-rays, produced inside the dump. The expected downstream leakage rate as a function of lead thickness is summarized in Table~\ref{tab:leakage}.
\begin{table}[h]
\centering
\caption{{Expected number of leakage photons per year reaching the downstream detector face as a function of lead beam-dump thickness, computed analytically using NIST XCOM mass attenuation coefficients for Pb and the full $\gamma$-$\gamma$ collider beam spectrum ($1.45\times10^{12}\,\mathrm{Hz}$, one year of operation), and confirmed by Geant4 simulation \cite{Allison:2016lfl,Allison:2006ve,GEANT4:2002zbu}. The adopted design thickness of $17\,\mathrm{cm}$ is highlighted in bold.}}

\label{tab:leakage}
\begin{tabular}{cc}
\hline
Pb thickness & Leakage photons/year \\
\hline
10\,cm & $1.3\times10^{7}$ \\
13\,cm & $3.9\times10^{3}$ \\
15\,cm & $20.6$ \\
\textbf{17\,cm} & $\mathbf{<0.1}$ \\
20\,cm & $3.2\times10^{-5}$ \\
\hline
\end{tabular}
\end{table}
As shown in Table~\ref{tab:leakage}, at $17\,\mathrm{cm}$ the expected leakage is reduced to fewer than $0.1$ photons per year, driven entirely by the hardest part of the beam spectrum ($>600\,\mathrm{keV}$); all photons below $600\,\mathrm{keV}$ are completely absorbed. The Geant4 simulation further confirms that zero secondary particles exit the downstream face of the dump—all backscattered and side-scattered radiation ($\sim10^{3}$ $\gamma$/year and $\sim70\,e^{-}$/year) exits the upstream or lateral surfaces and is therefore intercepted before reaching the detector. Cosmic-ray backgrounds are suppressed by anti-coincidence scintillator counters surrounding the photon detector; their contribution is estimated to be negligible compared to the $<0.1\,\mathrm{event/year}$ leakage budget. Prior to data-taking, the magnetic field $\vec{B}_{2}$ is removed to measure the ambient photon flux for background subtraction.

Downstream of the ALPs production region, the primary $\gamma$-rays are
removed by a dedicated beam dump system. This absorber eliminates the intense
photon flux before it reaches the wall, ensuring that the wall is not directly
irradiated by high-energy $\gamma$ rays. Consequently, potential ALPs production
from photon interactions in the wall material is strongly suppressed, and the
dominant ALPs source remains the well-defined magnetic conversion region.
An opaque wall is placed downstream of the beam dump to block residual or
scattered photons, while allowing ALPs to pass through essentially unimpeded.
Behind the wall, a second magnetic-field region, $|\vec{B}_{2}|=5\,\mathrm{T}$ and $L_2=0.2\,\mathrm{m}$, enables the
regeneration of photons via the inverse Primakoff effect. The regenerated
$\gamma$ rays are then detected by a high-resolution photon detector, which is
surrounded by anti-coincidence scintillator counters to suppress backgrounds
from cosmic rays and environmental radiation.

\section{Sensitivity to ALPs at the high-intensity \texorpdfstring{$\gamma-ray$}{gamma-ray} station}
\label{Sec::result}

In the high-intensity $\gamma$-ray station, $\gamma$-rays are emitted at a rate of 
$1.45\times10^{12}\,\mathrm{Hz}$, and their photon energy $\omega$ ranges from 
$0.02\,\mathrm{keV}$ to $720\,\mathrm{keV}$ \cite{zhou2025estimationinversecomptonscattering}. 
Only the photon field component parallel to the external magnetic field contributes to the photon-ALP conversion. In the $\gamma$-$\gamma$ collider, the polarization of the scattered photons is tunable via laser modulation \cite{DAngelo:2000rxv}. We can achieve a maximum of 90\% alignment of the scattered photon polarization parallel to the external magnetic field. As a conservative estimate, we assume that only half of the electric field components of photons are aligned with the external magnetic field and participate in the photon-ALPs conversion, so we add a factor of 1/2. We therefore define $n$ as the 
effective number of photons capable of interacting with the magnetic field. For one year of 
operation, this becomes
$n \;=\; \frac{1}{2}\,(1.45\times10^{12}\,\mathrm{s^{-1}})\,(3.15\times10^{7}\,\mathrm{s})
\;\approx\; 2.28\times10^{19}$.

 We assume that in our experimental setup, photons propagate in vacuum, there are no refractive effects, and all magnetic fields are uniform. Based on the model-independent ALP-photon coupling framework, the efficiency of converting photons to ALPs is given by \cite{Smarra:2024vzg}:

\begin{equation}
P_{\gamma \to a} = \frac{4 g_{a\gamma\gamma}^2 \omega^2 B_0^2}{ m_{a}^{4} + 4g_{a\gamma\gamma}^{2} \omega^{2} B_{0}^{2} } \sin^2 \left[ \frac{L_0q}{2}\right],
\end{equation}

\begin{equation}
q =  \sqrt{ \omega^2 + \frac{1}{2} \left( \sqrt{m_a^4 + 4g_{a\gamma\gamma}^2 \omega^2 B_0^2} - m_a^2 \right) } - \sqrt{ \omega^2 - \frac{1}{2} \left( \sqrt{m_a^4 + 4g_{a\gamma\gamma}^2 \omega^2 B_0^2} + m_a^2 \right) },
\end{equation}

where $B_0=|\vec{B_0}|$ is the magnetic field strength in the collider, $L_0$ the interaction length in $\vec{B_0}$. $m_{a}$ is the mass of an ALP, and $g_{a\gamma \gamma}$ is the photon-ALP coupling constant.
 
We assume $m_{a} \ll \omega$ and $4g_{a\gamma\gamma}^2 \omega^2 B_0^2 \ll m_a^4$, so we have $q \approx \frac{m_{a}^{2}}{2 \omega}$ and $\frac{\omega}{\sqrt{\omega^2 - m_a^2}} \simeq 1$. The efficiency of converting photons to ALPs is transformed into the formula we are familiar with \cite{Capparelli:2015mxa}:

\begin{equation}
P_{\gamma \to a} = g_{a\gamma\gamma}^2 B_{\rm{0}}^2 \frac{\omega}{\sqrt{\omega^2 - m_a^2}} \left( \frac{\sin^2(q L_{\rm{0}}/2)}{q^2}\right).
\end{equation}

So the efficiency of converting photons to ALPs is:

\begin{equation}
P_{\gamma \to a} = 4g_{a\gamma\gamma}^2 B_{\rm{0}}^2\frac{\omega^{2}}{m_{a}^{4}}\sin^2(\frac{m_{a}^{2}L_{\rm{0}}}{4\omega} ).
\label{eq:efficiency1}
\end{equation}

If  $m_{a}^{2}L_{\rm{0}} \ll 4\omega$, the efficiency of converting photons to ALPs is transformed into \cite{Cameron:1993mr}:

\begin{equation}
P_{\gamma \to a} =\frac{L_{\rm{0}}^{2}B_{\rm{0}} ^{2} g_{a \gamma \gamma } ^{2}}{4}.
\label{eq:a_efficiency}
\end{equation}

Similarly, $P_{a \to \gamma}$ is:
\begin{equation}
P_{a \to \gamma} = 4g_{a\gamma\gamma}^2 B_{\rm{2}}^2\frac{\omega^{2}}{m_{a}^{4}}\sin^2(\frac{m_{a}^{2}L_{\rm{2}}}{4\omega} ),
\label{eq:efficiency2}
\end{equation}

where $B_2=|\vec{B_2}|$ is the applied magnetic field strength, and $L_2$ is the length of the magnetic field to regenerate photons.

When the mass of the ALPs approaches the photon energy, the previously mentioned approximation will fail, we should consider Eq.~\ref{eq:efficiency1} and Eq.~\ref{eq:efficiency2}. So, the expected regenerated photon count $R$ is:

\begin{equation}
R = \int P_{\gamma \to a}\left(\omega\right)P_{a \to \gamma}\left(\omega\right)\eta\, dn,
\label{eq:count rate}
\end{equation}
where $\rm{\eta}$ is the efficiency of the photon detector.

\begin{figure}
    \centering
    \includegraphics[width=\linewidth]{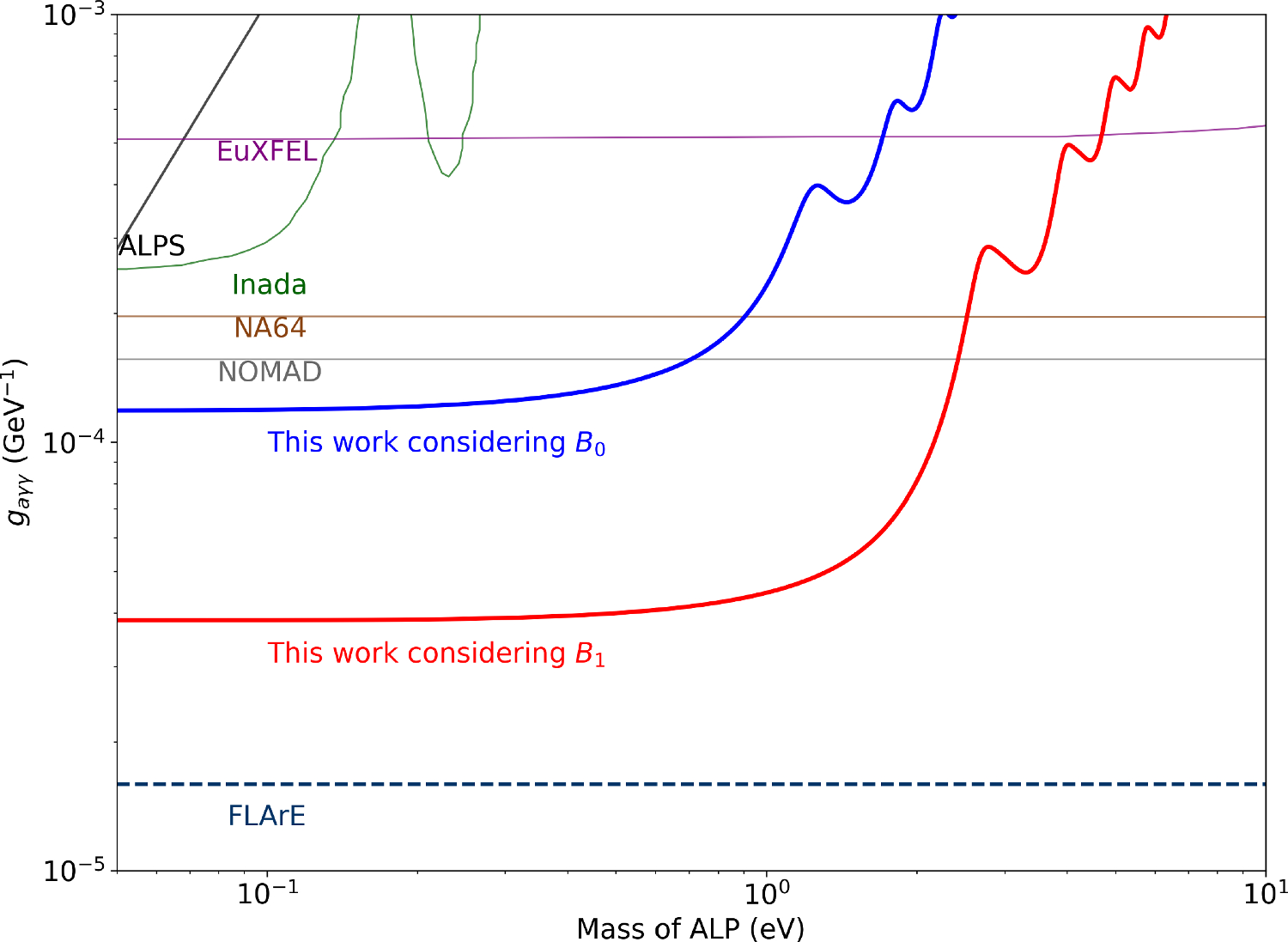}
    \caption{Comparison of the projected sensitivity of our setup (red and blue) with other LSW experiments (LSW experiments use a 95\% confidence level based on Poisson statistics), including EuXFEL \cite{Halliday:2024lca},     
Inada \cite{Inada:2016jzh},     
NA64 \cite{NA64:2020qwq},     
ALPS \cite{Ehret:2010mh},     
and NOMAD~\cite{NOMAD:2000usb}. The dashed line indicates the projected sensitivity for FLArE, based on a proposal \cite{Kling:2022ehv}. In contrast to EuXFEL \cite{Halliday:2024lca}, which uses electric fields for ALP-photon conversion, other LSW experiments employ magnetic fields. The Inada dataset was directly reproduced from the literature \cite{Inada:2016jzh}, while the other experimental datasets were obtained from the GitHub website (\url{https://github.com/cajohare/AxionLimits/tree/master/limit_data/AxionPhoton}).    
The improvement, particularly in the sub-eV to eV mass range, arises mainly from the substantially enhanced photon energy provided by the $\gamma$-ray source. The sensitivity that uses only deflection magnetic field $\vec{B_0}$ is shown in blue, and the sensitivity that adds an additional magnetic field $\vec{B_1}$ inside the $\gamma$-$\gamma$ collider is shown in red.}
    \label{fig:gagamma}
\end{figure}

If no regenerated photon signal is detected, the corresponding parameter region can be excluded at the 95\% confidence level based on standard Poisson statistics. In the background-free case, this criterion corresponds to a threshold of $R_{\rm th} = 3.0$ expected signal events, obtained from the condition $1 - e^{-\mu} = 0.95$. Assuming $\eta=100\%$, the projected sensitivity obtained under this protocol is shown in
Fig.~\ref{fig:gagamma}, which presents the expected exclusion boundary in the $g_{a\gamma\gamma}$–-$m_a$ plane over the mass range $5\times10^{-2}\text{--}10^{1}\,\mathrm{eV}$.To provide a clear context for these results, the key experimental parameters and the provenance of the comparison curves shown in Fig.~\ref{fig:gagamma} are summarized in Table~\ref{tab:compare}. Compared with previous LSW experiments, our setup provides a significant improvement, particularly in the sub-eV to eV mass range.

The background-free assumption is justified by the Geant4 simulation described in Sec.~\ref{Sec::method}: a $17\,\mathrm{cm}$ lead beam dump suppresses the leakage photon flux to fewer than $0.1$ events per year of operation. With a leakage background $b < 0.1$, the Poisson exclusion threshold shifts from $3.0$ to $3.0 + b < 3.1$ expected signal events, a change of less than $3\%$ that has negligible impact on the quoted sensitivity. Regarding detector efficiency: the expected signal count scales as $R \propto g_{a\gamma\gamma}^4\,\eta$, so the sensitivity on the coupling scales as $g_{a\gamma\gamma} \propto \eta^{-1/4}$. For a realistic detector efficiency of $\eta = 50\%$, the coupling reach degrades by a factor of $2^{1/4} \approx 1.19$, weakening the sensitivity from $3.84\times10^{-5}$ to $\sim4.57\times10^{-5}\,\mathrm{GeV^{-1}}$—a modest reduction that preserves the improvement over all previous laboratory LSW experiments.

With an effective photon yield of $n_{\rm eff}=2.28\times 10^{19}$ after one year of operation, the theoretical sensitivity reaches $g_{a\gamma\gamma} = 1.19\times 10^{-4}\,\mathrm{GeV^{-1}}$ at ALP masses below $0.1\,\mathrm{eV}$ when only considering the deflection magnetic field $\vec{B}_{0}$ of the collider, surpassing the performance of all previous laboratory-based searches. Adding a magnetic field $\vec{B}_{1}$ (identical to $\vec{B}_{2}$) in front of the beam dump system, the sensitivity can be much improved to $g_{a\gamma\gamma} = 3.84\times 10^{-5}\,\mathrm{GeV^{-1}}$ at ALP masses below $0.1\,\mathrm{eV}$. Even if the residual background is included conservatively, the threshold changes only from 3.0 to $R < 3.1$, which modifies the coupling reach by less than 1\%. Therefore, the quoted sensitivity is essentially unchanged. At present, the best limit provided by the LSW method is from NOMAD~\cite{NOMAD:2000usb}, which reported $g_{a\gamma\gamma} \leq 1.5\times 10^{-4}\,\mathrm{GeV^{-1}}$ under $m_a$ less than $1\,\mathrm{eV}$. If this sensitivity is taken into account, our system is expected to detect 710 signals.

\begin{table}[h]
\centering
\caption{Key parameters of the ALP searches compared in Fig.~\ref{fig:gagamma}. The table details the experimental methods, characteristic photon energy scales, and whether the displayed curves represent existing exclusion bounds or future projections. All external limits and projections are adopted directly from the cited references for benchmarking purposes.}
\label{tab:compare}

\begin{tabular}{@{}lllll@{}}
\toprule
Experiment & Experimental method & Photon-energy scale & Result type & Curve source \\ \midrule
EuXFEL     & Crystal-based LSW   & $9.8\,\mathrm{keV}$ & Bound       & Ref.~\cite{Halliday:2024lca} \\
Inada      & Magnetic LSW        & $9.5\,\mathrm{keV}$ & Bound       & Ref.~\cite{Inada:2016jzh} \\
NA64       & Beam-based ALP search & $\ge 15\,\mathrm{GeV}$ & Bound     & Ref.~\cite{NA64:2020qwq} \\
ALPS       & Magnetic LSW        & $2.33\,\mathrm{eV}$ & Bound       & Ref.~\cite{Ehret:2010mh} \\
NOMAD      & Beam-based ALP search & $5\,\mathrm{GeV}$--$140\,\mathrm{GeV}$ & Bound & Ref.~\cite{NOMAD:2000usb} \\
FLArE      & Proposed beam-based search & $1\,\mathrm{GeV}$--$10\,\mathrm{TeV}$ & Projection & Ref.~\cite{Kling:2022ehv} \\ \midrule
This work  & $\gamma$-ray LSW at magnetic field & $0.02\,\mathrm{keV}$--$720\,\mathrm{keV}$ & Projection & This work \\ \bottomrule
\end{tabular}
\end{table}

\section{Summary}
\label{Sec::Summary}
We have investigated a practical extension of the light-shining-through-walls technique that makes direct use of the high-intensity $\gamma$-ray beam naturally provided by a $\gamma$-$\gamma$ collider. The energetic and well-collimated photons produced via inverse Compton scattering offer an efficient ALPs production stage, while a downstream regeneration region completes the LSW setup without requiring significant modifications to the collider infrastructure. Under conservative assumptions, one year of operation yields a projected sensitivity of $g_{a\gamma\gamma} \simeq 3.84\times10^{-5}\,\mathrm{GeV^{-1}}$ for $m_a \lesssim 0.1\,\mathrm{eV}$ when an additional magnetic region is included upstream of the beam dump. Our setup improves existing laboratory limits by up to a factor of 4.
Although this sensitivity is still above the QCD axion band, it represents a practical laboratory step toward closing the large gap between current LSW limits at the $10^{-4}\mathrm{GeV}^{-1}$ level and the $10^{-9}\mathrm{GeV}^{-1}$ QCD axion target near the eV scale.

Because the method relies solely on controlled laboratory conditions, it avoids the astrophysical uncertainties affecting stellar bounds. These high-energy photons probe the sub-eV to eV mass range, providing a crucial model-independent cross-check for helioscope and astrophysical limits. Although the projected sensitivity already exceeds existing LSW exclusion limits, there remains significant room for improvement. In the low-mass region, the regenerated photon rate scales as $R \propto g_{a\gamma\gamma}^4 (BL)^4$. Modest increases in magnetic field strength or magnet length can translate into a significant enhancement of the detection capability, making this proposal not only an independent experimental measurement, but also a scalable pathfinder for next-generation collider-based LSW experiments.

\backmatter

\bmhead{Acknowledgements}
 The authors thank Mr. Junxi Liu, Drs. Meiyu Si and Yongsheng Huang for the useful discussions. This work is supported in part by the Guangdong Provincial Key Laboratory of Advanced Particle Detection Technology (2024B1212010005), the Guangdong Provincial Key Laboratory of Gamma-Gamma Collider and Its Comprehensive Applications (2024KSYS001), the Fundamental Research Funds for the Central Universities, and the Sun Yat-sen University Science Foundation.

\section*{Declarations}

\begin{itemize}
\item Funding

Not applicable.

\item Competing interests

No competing interests.

\item Ethics approval and consent to participate

Not applicable.

\item Consent for publication

Not applicable.

\item Data availability 

Not applicable.

\item Materials availability

Not applicable.

\item Code availability 

Not applicable.

\item Author contribution

Jie Feng conceived the idea. Zi-Yao Yan conducted data analysis. Both authors discussed the results and contributed to the final manuscript.

\end{itemize}

\noindent

\bibliography{sn-bibliography}

\end{document}